\documentclass[aps,prl,superscriptaddress,twocolumn,showpacs]{revtex4-1}

\usepackage{amssymb}
\usepackage{graphicx}
\usepackage{bm}
\usepackage{color}
\usepackage{longtable}
\usepackage{amsmath}
\usepackage{tabularx}
\usepackage{placeins}
\usepackage{afterpage}
\usepackage[colorlinks,
linkcolor=blue,
citecolor=blue,
urlcolor=blue]{hyperref}

\begin{document}
\title{Injection dynamics of direct-laser accelerated electrons in a relativistic transparency regime}
\author{K. Jiang}
\affiliation{Graduate School, China Academy of Engineering Physics, Beijing 100088,
People's Republic of China}
\affiliation{Institute of Applied Physics and Computational Mathematics,
Beijing 100094, People's Republic of China}
\author{C. T. Zhou} \email{zcangtao@iapcm.ac.cn}
\affiliation{Institute of Applied Physics and Computational Mathematics, Beijing 100094, People's Republic of China}
\affiliation{Center for Advanced Material Diagnostic Technology, Shenzhen Technology University, Shenzhen 518118, People's Republic of China}
\affiliation{College of Optoelectronic Engineering, Shenzhen University, Shenzhen 518060, People's Republic of China}
\affiliation{HEDPS, Center for Applied Physics and Technology and School of Physics, Peking University, Beijing 100871, People's Republic of China}
\author{T. W. Huang} \email{taiwu.huang@szu.edu.cn}
\affiliation{College of Optoelectronic Engineering, Shenzhen University, Shenzhen 518060, People's Republic of China}
\author{L. B. Ju}
\affiliation{Graduate School, China Academy of Engineering Physics, Beijing 100088, People's Republic of China}
\affiliation{Institute of Applied Physics and Computational Mathematics,
Beijing 100094, People's Republic of China}
\author{H. Zhang}
\affiliation{Institute of Applied Physics and Computational Mathematics,
Beijing 100094, People's Republic of China}
\author{B. Qiao}
\affiliation{HEDPS, Center for
Applied Physics and Technology and School of Physics, Peking University, Beijing 100871, People's Republic of China}
\author{S. C. Ruan}
\affiliation{Center for Advanced Material Diagnostic Technology, Shenzhen Technology University, Shenzhen 518118, People's Republic of China}
\affiliation{College of Optoelectronic Engineering, Shenzhen University, Shenzhen 518060, People's Republic of China}

\date{\today }

\begin{abstract}
The dynamics of electron injection in the direct laser acceleration (DLA) regime was investigated by means of
three-dimensional particle-in-cell simulations and theoretical analysis. It is shown that when
an ultra-intense laser pulse propagates into a near-critical density or relativistically transparent plasma,
the longitudinal charge-separation electric field excites an ion wave. The ion wave modulates the local
electric field and acts as a set of potential wells to guide the electrons, located on the edge of the plasma
channel, to the central region, where the DLA takes place later on. In addition, it is pointed out that the
self-generated azimuthal magnetic fields tend to suppress the injection process of electrons by deflecting
them away from the laser field region. Understanding these physical processes paves the way for further
optimizing the properties of direct-laser accelerated electron beams and the associated X/gamma-ray sources.
\end{abstract}

\maketitle


High-energy electron beams are attractive for many applications ranging from fast ignition of
inertial confinement fusion \cite{tab}, radiography \cite{man}, and novel light sources \cite{kne,giu,hua1},
to neutron sources \cite{pom}. With the ability of supporting huge accelerating
electric fields (above $100$ $\mathrm{GV/m}$), the laser-based plasma accelerators, which are promising
to revolutionize the conventional accelerator technologies, recently have attracted much research
attention. Based on laser-plasma interactions, two distinct regimes have been proposed to produce
high-energy electron beams, which are known as the laser wake-field acceleration (LWFA) and the direct
laser acceleration (DLA). In LWFA regime, so far quasi-monoenergetic electron beams with energies up to
several $\mathrm{GeV}$ have been obtained \cite{taj,lee,kim,wan,liu0}. 
However, the rather low-density plasma (typically less than $10^{20}$ $\mathrm{cm^{-3}}$) used in this scheme
generally limits the charges of accelerated electrons to about $100$ $\mathrm{pC}$, which makes
this regime infeasible for many applications that requires high-charge electron beams.

The DLA regime becomes dominant for electron acceleration in relatively high density plasmas,
and in this regime high-charge (higher than $10$ $\mathrm{nC}$) electron beams with energies
up to hundreds of $\mathrm{MeV}$ have been obtained \cite{man1}. This regime occurs when the
betatron oscillation frequency of the electrons in the laser-produced plasma channel is equal
to the Doppler-shifted laser frequency \cite{puk1}. Recently, several different schemes have
been proposed to enhance the electron energy in this regime \cite{are,rob,jia}. So far little
attention has been paid to the injection process of the electrons in this regime \cite{are2,hua2},
which plays a significant role on the subsequent electron acceleration process and
determines the qualities of the accelerated electrons. In particular, the DLA electron beams usually display
as broad energy spectra. Understanding the injection mechanism is an important step to optimize and improve
the electron beam qualities in this regime. The injection process of electrons in the plasma
channel seems complicated and chaotic \cite{man1,hu}, and only few works discuss about it. For a relatively
low-density plasma, the laser field could excite surface plasma wakes along the plasma channel, which traps
and pre-accelerates the electrons so that they can enter into the phase of direct-laser acceleration \cite{nas} .
However, in a relativistic transparency regime with near-critical density plasma, the injection dynamics of
DLA electrons is not yet clear. In particular, in this case the motion of the background ions, which can
be dragged by the large charge-separation electric fields, cannot be neglected as done in previous works.
\begin{figure*}
  \centering
  \includegraphics[width=17cm]{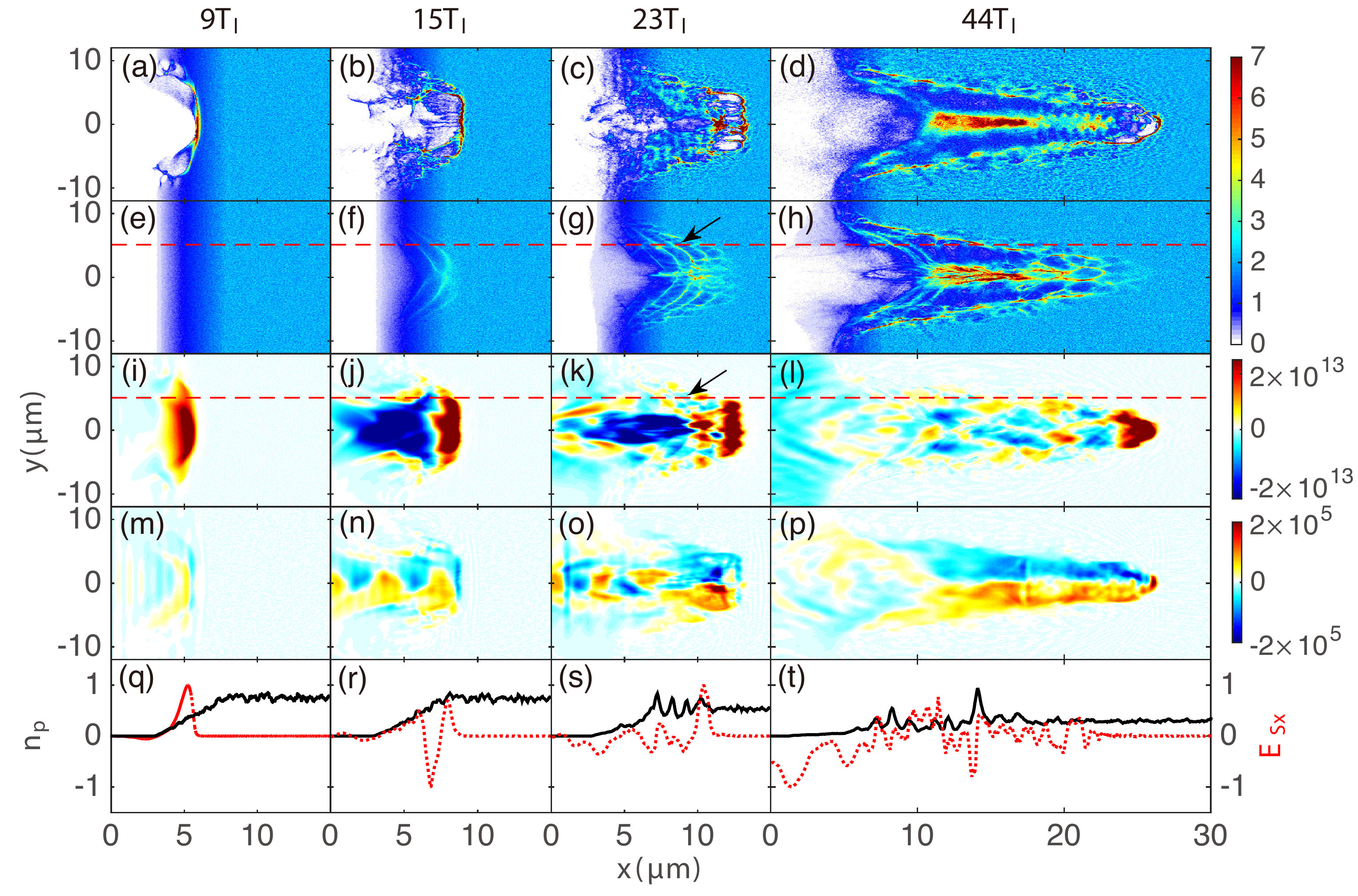}\\
  \caption{Results of the 3D-PIC simulation; Results are shown as cuts in $(x,y)$ plane at $z=0$ $\mathrm{\mu m}$ in the first four rows and line graphs at $y=5$ $\mathrm{\mu m}$, $z=0$ $\mathrm{\mu m}$ (marked by red dash lines in (e)-(l)) in the last row. From left to right, results at different times: $t=9T_l$, $15T_l$, $23T_l$, $44T_l$, respectively. From top to bottom, distributions of electron density ($n_e/n_c$), ion density ($n_p/n_c$), longitudinal charge-separation electric field $E_{Sx}$ ($\mathrm{V/m}$), azimuthal magnetic field $B_{S\theta}$ ($\mathrm{T}$), and normalized ion density profile $n_i/max(n_i)$ (black solid line) with normalized longitudinal charge-separation electric field profile $E_{Sx}/max(\left | E_{Sx} \right |)$ (red dot line), respectively. The black arrows in (g) and (k) point at the ion wave and corresponding localized electric field $E_{osc}$.}\label{fig_1}
\end{figure*}

In this paper, we present the first investigation on the injection dynamics of direct-laser accelerated
electrons in the relativistic transparency regime. It is shown that the ion wave, excited by the longitudinal
charge-separation electric field, plays a crucial role on the electron injection process. The localized electric
field induced by the ion motion, acts as a set of potential wells, which trap and guide the electrons
from the edge of the plasma channel into the central region so that they can be directly accelerated by the laser
field. In addition, with the increase of the self-generated azimuthal magnetic field, the injected electrons are
deflected away from the laser field region and thus the injection process of electrons is gradually terminated.

To investigate the electron injection mechanism in DLA regime, three-dimensional (3D) particle-in-cell (PIC) simulations were performed with EPOCH code \cite{arb}. In the simulations, a right-hand circularly polarized (CP) Gaussian laser pulse with central wavelength $\lambda =800$ $\mathrm{nm}$, laser period $T_l=\lambda /c$, peak intensity $I_0=2 \times 10^{21}$ $\mathrm{W/cm^{2}}$, and spot radius $r_0=5$ $\mathrm{\mu m}$ is normally incident from the left boundary ($x=0$ $\mathrm{\mu m}$). The duration of the laser pulse is $21T_l$ with a rising time of $2T_l$. The plasma target consisting of electrons and protons is placed in the region of $3$ $\mathrm{\mu m} <x<32$ $\mathrm{\mu m}$. In the laser propagation direction, the plasma density increases linearly form $0$ to $n_0=2n_c$ in a distance of $5$ $\mathrm{\mu m}$, and then remains constant. Here $n_c=m_e\omega_0^{2}\epsilon _0/e^2$ is the critical plasma density, where $m_e$ is the rest electron mass, $\omega_0$ is the laser frequency, $e$ is the elementary charge, and $\epsilon _0$ is the vacuum permittivity. In the radial direction, the plasma density is uniform. The initial electron temperature $T_e$ is $1$ $\mathrm{keV}$, and proton temperature $T_i$ is $100$ $\mathrm{eV}$. The simulation box is $32 \times 24 \times 24$ $\mathrm{\mu m^{3}}$ with a grid of $600 \times 450 \times 450$ and 8 particles per cell. In the present setup, the plasma is relativistically transparent for the ultra-intense laser pulse and DLA becomes the dominant regime for producing high-current high-energy electron beam \cite{hua,ju}.

The corresponding simulation results are shown in Fig. {\ref{fig_1}}. At $t=9T_l$, the laser pulse impinges on the plasma and its ponderomotive force expels the electrons from the central region. A typical dense electron layer is formed in front of the laser pulse, as shown in Fig. \ref{fig_1}(a). As the response time for ions is relatively long, the background ions are almost immobile at this moment. In this case, the longitudinal charge-separation electric field ($E_{back}$) increases linearly with $x$, as shown in Fig. \ref{fig_1}(q), which is similar with the field generated at the front surface of an overdense plasma \cite{mac}. At $t=15T_l$, the longitudinal charge-separation field starts to drag the background ions. Fig. \ref{fig_1}(f) shows that a linear ion wave with central wavelength $\lambda_i \sim 1.12$ $\mu m$ and ion velocity $v_i \sim 0.11c$ is excited, where $c$ is the speed of light in vacuum. Such ion wave is also observed in the laser-driven solitary ion acceleration \cite{zha}. The ion wave can induce a localized electric field $E_{osc}$, which alternates its sign with zero points located at the peaks of ion wave \cite{liu}. In this case, the charge-separation electric field can be written as $E_{Sx}=E_{back}+E_{osc}$ and it becomes nonlinear with a large gradient at the point $E_{Sx}=0$, as show in Fig. \ref{fig_1}(r). At a later moment $t=23T_l$, the ion wave is left behind by the laser pulse as the ion velocity is much smaller than the laser group velocity. The electric field $E_{osc}$ generated by the ion wave, which is earlier overlapped by the electric field $E_{back}$, now becomes isolated with an amplitude of $10^{12}$ $\mathrm{V/m}$, as can be seen in Figs. \ref{fig_1}(k) and (s). This electric field links the channel edge to the channel axis and acts as a set of potential wells. Electrons from the channel edge will move towards the central region following the path where $E_{osc}=0$, as a result, a dense electron bunch is formed in the plasma channel, as can be seen in Fig. \ref{fig_1}(c). In contrast, when the ions are fixed as background in the simulation (not shown here), the generated dense electron bunch is distributed around the front surface of the target ($x=3$ $\mathrm{\mu m}$). At $t=44T_l$, it is shown from Fig. \ref{fig_1}(d) that a regularly modulated energetic electron bunch with density above $3n_c$ is formed in the central region, which is indicative of DLA \cite{hua}. In addition, it is shown from Figs. \ref{fig_1}(d) and (h) that at the tail of the electron bunch, several electron density spikes are overlapped with the ion density spikes, which connects the channel edge with the channel axis. This indicates that these electrons are injected along the ion density spikes, which correspond to the electric field of $E_{osc}=0$. Fig. \ref{fig_1}(p) shows that intense azimuthal magnetic field $B_{S\theta}$ is induced by the forward-moving energetic electron beam in the plasma channel shown in Fig. \ref{fig_1}(d). The strength of the magnetic field increases with time and gradually becomes saturated at a value about $2\times 10^5$ $\mathrm{T}$, as shown in Figs. \ref{fig_1}(m), (n), (o), and (p).

\begin{figure}
  \centering
   \includegraphics[width=8.5cm]{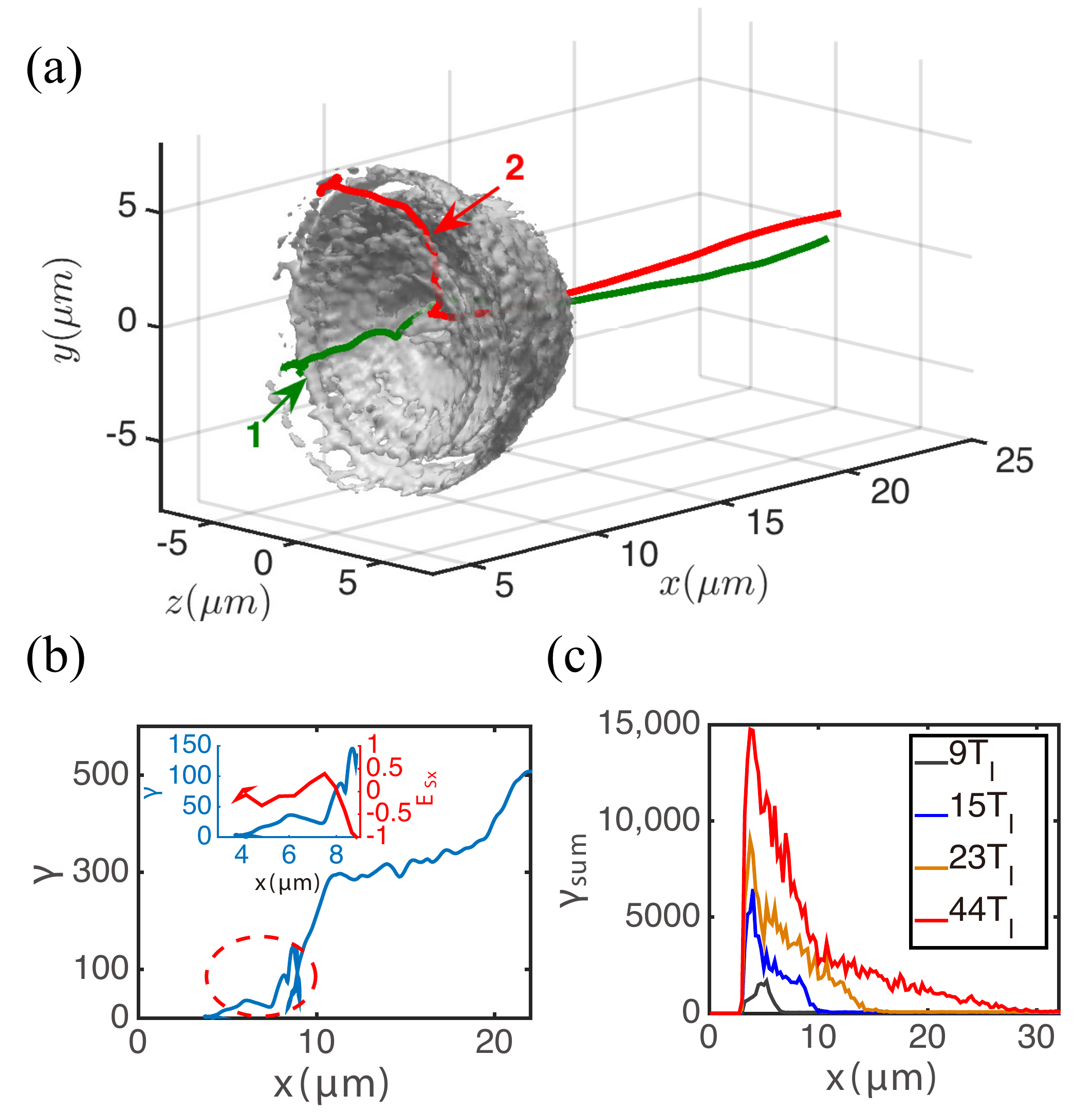}\\
 \caption{Results of particle tracking: (a) trajectories of sample electrons and the isosurface distribution of ion density, where the isovalue is set as $2.28n_c$. And the ion density is cut at $t=23T_l$. The color lines marked by numbers correspond to trajectories of individual electrons; (b) The partial evolution of the Lorentz factor $\gamma$ for electron number 2; (c) The sum of Lorentz factor $\gamma$ at $t=9T_l$, $15T_l$, $23T_l$ and $44T_l$ for counted electrons from different initial positions. The inset in (b) shows the local partial evolution of the Lorentz factor $\gamma$ and the normalized total longitudinal quasi-static electric field $E_{Sx}/max(\left | E_{Sx} \right |)$ felt by electron number 2.}\label{fig_2}
\end{figure}
Particle tracking method is also employed to give an insight into the electron injection process. Fig. \ref{fig_2}(a) shows the 3D isosurface distribution of ion density and the typical trajectories of high-energy electrons. It is shown that ions are distributed as a set of funnel-shaped layers, which correspond to the ion wave discussed above. After an initial injection process, the electrons enter into the stage of DLA and experience betatron oscillations in the plasma channel, as shown in Fig. \ref{fig_2}(a). The corresponding Lorentz factor $\gamma$ of the DLA electron can exceed 500, as can be seen in Fig. \ref{fig_2}(b). The two sample electrons marked with different numbers, shown in Fig. \ref{fig_2}(a), also represent different groups of electrons that undergo different injection processes. The electron 1, which is initially located at the center of plasma channel, is not expelled by the ponderomotive force and performs betatron oscillations at an early time. The electron 2 is initially located at the edge of the channel and needs to be injected into the central region to enter into the DLA stage. Fig. \ref{fig_2}(a) shows that the electron 2 moves to the central region along the surface of a certain ion layer during the injection process. In addition, Fig. \ref{fig_2}(b) shows that the electric field experienced by the electron 2 is approximately zero during the initial injection process, which indicates that the electron moves along the peak of ion wave with $E_{osc}=0$. For this group of electrons, as the work done by the radial charge-separation field $E_{Sr}$ is positive, they can be pre-accelerated during the injection process, with the Lorentz factor $\gamma$ increased above 20, as shown in Fig. \ref{fig_2}(b). It is found that almost all the resonance electrons that are initially located at the channel edge, experience similar injection dynamics. Furthermore, the simulations with linearly polarized laser pulse are also carried out and it is shown that the electrons also show similar injection dynamics.

In addition, it is noted that some injected electrons experience a special juncture before DLA, where they are exerted by a negative longitudinal electric field $E_{Sx}$, as can be seen in the inset of Fig. \ref{fig_2}(b). This field plays a crucial role in reducing the dephasing rate and leads to the generation of superponderomotive electrons \cite{rob}. After the injection process, the electrons are confined in the plasma channel by the self-generated azimuthal magnetic field $B_{S\theta}$ and the radial charge-separation field $E_{Sr}$. They can be directly accelerated by the laser field when the betatron frequency is close to the Doppler-shifted laser frequency. Fig. \ref{fig_2}(c) shows the sum of the Lorentz factor ($\gamma_{sum}$) of the counted electrons as a function of their initial positions. It is shown that there is a sudden change in the gradient of $\gamma_{sum}$ around $x=10$ $\mathrm{\mu m}$, which also corresponds to the injection range of electrons shown in Figs. \ref{fig_2}(a) and (b). This indicates that most of the resonance electrons comes from the front surface of the target and the injection process is nearly prohibited for the electrons from the middle area of the target.

The ion wave is induced by the longitudinal charge-separation electric field $E_{back}$, which is commonly observed in near-critical density or relativistically transparent plasmas. By using the ion fluid model and switching to the wavefront coordinate, the periodic electric field $E_{osc}$, induced by the ion wave, can be given analytically as  $\tilde{E}_{osc}=\chi^{1/3}-\frac{2(1-\beta_0)}{\chi^{1/3}}-\frac{3\xi}{3-2\beta_0}$, where $\chi=3\xi+\sqrt{9\xi^2+8(1-\beta_0)^3}$, $\beta_0=\beta_i(\xi = 0)$, $\beta_i=v_i/v_{ph}$, $\xi=\omega_i(x/v_{ph}-t)$, $\omega_i=\sqrt{q_ien_0/m_i\epsilon_0}$, $v_i=dx/dt$ is the ion velocity along the longitudinal direction, $v_{ph}$ is the phase velocity of the wake, $m_i$ is the ion mass, $q_i$ is the ion charge (for hydrogen plasma $q_{i}=1$), and $n_i$ is the ion density \cite{liu}. Thus in the SI units, one has $E_{osc}=\frac{v_{ph}m_i\omega _i}{e}\left [\chi ^{1/3}-\frac{2(1-\beta_0)}{\chi^{1/3}}-\frac{3\omega_i\xi_r/v_{ph}}{3-2\beta_0} \right ]$, where $\xi_r$ is the distance to the peak of ion wave. Based on these equations, one can examine the dynamics of a single electron in the presence of periodic electric field $E_{osc}$, the radial charge-separation field $E_{Sr}$, and the self-generated azimuthal magnetic field $B_{S\theta}$. For simplicity, the electron dynamics in a two dimensional plane $(x,y)$ is considered. In this case, $E_{Sr}=E_{Sy}=(\frac{m_ec^2}{er_0})a_0^2\frac{y}{r_0}e^{\frac{y^2}{r_0^2}}(1+a_0^2e^{-\frac{y^2}{r_0^2}})^{-\frac{1}{2}}$, where $a_0$ is the normalized laser electric field and $r_0$ is the laser spot radius. The azimuthal magnetic field  $B_{S\theta}=B_{Sz}=-B_0\frac{y}{r_0}e^{-\frac{y^2}{r_0^2}}$, where $B_0$ is the amplitude of the magnetic field and can be obtained from the simulation results \cite{tsa,qia}. Then the equations of motion for the test electron can be written as
\begin{equation}
\frac{dp_x}{dt}= -eE_{osc}-ev_yB_{Sz}
\end{equation}
\begin{equation}
\frac{dp_y}{dt}=-eE_{Sy}+ev_xB_{Sz}
\end{equation}
\begin{equation}
m_ec^2\frac{d\gamma}{dt}=-ev_xE_{osc}-ev_yE_{Sy}
\end{equation}
After some algebra, one could obtain
\begin{equation}
\frac{d\xi_r}{dt}=v_x-v_i
\end{equation}
\begin{equation}
\begin{aligned}
\frac{dv_x}{dt}=&\ \sqrt{1-\frac{v_x^2+v_y^2}{c^2}} \{ \left( \frac{v_x^2}{c^2}-1 \right)\frac{m_i}{m_e}v_{ph}\omega_i \\ 
&\ \left[ \chi^{1/3}-\frac{2(1-\beta_0)}{\chi^{1/3}}-\frac{3\omega_i\xi_r/v_{ph}}{3-2\beta_0} \right]+v_xv_y\frac{a_0^2}{r_0^2}ye^{-\frac{y^2}{r_0^2}}\\
&\ \left( 1+a_0^2e^{-\frac{y^2}{r_0^2}} \right)^{-1/2} +\frac{e}{m_e}v_yB_0\frac{y}{r_0}e^{-\frac{y^2}{r_0^2}} \}
\end{aligned}
\end{equation}
\begin{equation}
\frac{dy}{dt}=v_y
\end{equation}
\begin{equation}
\begin{aligned}
\frac{dv_y}{dt}=&\ \sqrt{1-\frac{v_x^2+v_y^2}{c^2}} \{ \frac{m_i}{m_e}\frac{v_xv_y}{c^2}v_{ph}\omega_i \\
&\ \left[ \chi^{1/3}-\frac{2(1-\beta_0)}{\chi^{1/3}}-\frac{3\omega_i\xi_r/v_{ph}}{3-2\beta_0} \right] \\
&\ + (v_y^2-c^2)\frac{a_0^2}{r_0^2}ye^{-\frac{y^2}{r_0^2}} \left( 1+a_0^2e^{-\frac{y^2}{r_0^2}} \right)^{-1/2} \\
&\ -\frac{ev_xB_0}{m_e}\frac{y}{r_0}e^{-\frac{y^2}{r_0^2}} \}
\end{aligned}
\end{equation}

\begin{figure}
  \centering
  \includegraphics[width=8.5cm]{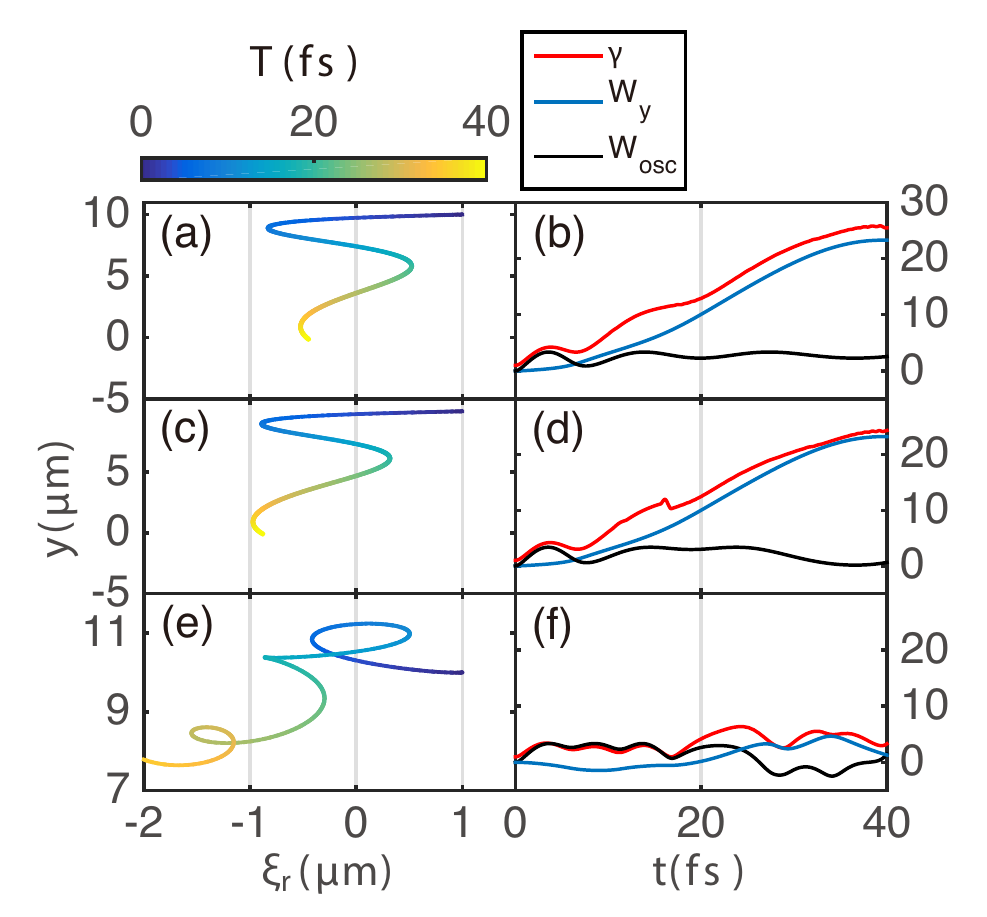}\\
  \caption{Results of single-particle simulations based on Eqs.(4)-(7) with initial conditions at $t=0$, $\xi_{r0}=1$ $\mathrm{\mu m}$, $y_0=10$ $\mathrm{\mu m}$, $v_{x0}=v_{y0}=0$, $v_{ph}=0.883c$, $\beta_0=0.1557$, $a_0=27$, $r_0=5$ $\mathrm{\mu m}$, total iteration time $t_{end}=40$ $\mathrm{fs}$ and the iterative time span $0.001$ $\mathrm{fs}$. From up to bottom, results at different $B_{Sz}$: (a) and (b) $0$ $\mathrm{T}$, (c) and (d) $1$ $\times 10^4 \mathrm{T}$, (e) and (f) $2 \times 10^5$ $\mathrm{T}$. From left to right, electron trajectories and the time evolution of the electron Lorentz factor $\gamma$, the dimensionless energy gain from the radial electric field $W_y=-\int_{0}^{t}ev_{y}E_{Sy}dt$ and the longitudinal periodic electric field $W_{osc}=-\int_{0}^{t}ev_{x}E_{osc}dt$.}\label{fig_3}
\end{figure}

The above equations can well describe the injection dynamics of electrons before entering into the DLA stage. However, the analytic solution is difficult to obtain due to the nonlinearity. It is shown from Eqs.(5) and (7) that the azimuthal magnetic field has a great influence on the electron injection dynamics. The azimuthal magnetic field $B_{Sz}$ is generated by the forward-moving electrons on the axis and increases from $0$ to $2\times 10^5$ $\mathrm{T}$ during the electron acceleration process. Thus one could vary $B_0$ in Eqs.(5) and (7) to represent different stages. The numerical results are given in Fig. {\ref{fig_3}} with $B_0=0$ $\mathrm{T}$, $1\times 10^4$ $\mathrm{T}$, and $2\times 10^5$ $\mathrm{T}$, respectively. Figs. {\ref{fig_3}}(a) and (b) show the initial moment without stable electron current inside the plasma channel. The electron moves towards the channel axis and oscillates around the peak of ion wave ($\xi_r=0$ $\mathrm{\mu m}$). This indicates that the electron is injected towards the central region along the peak of ion wave, and during the injection process, the electron is pre-accelerated by the radial electric field and its Lorentz factor increases to around 25, as shown in Fig. {\ref{fig_3}}(b), which is in good agreement with the simulation result shown in Fig. \ref{fig_2}(b). Figs. {\ref{fig_3}}(c) and (d) show the stage that the injection process is not saturated. In this case, the electron shows similar dynamics with that in Figs. {\ref{fig_3}}(a) and (b), though it is deflected a bit towards the negative direction of $\xi_r$ by the azimuthal magnetic field. Figs. {\ref{fig_3}}(e) and (f) show the saturate stage of the electron injection process, in which the electron is deflected largely to the negative direction of $\xi_r$ and it cannot reach the central region of the plasma channel. In this case, the pre-acceleration of the electron is negligible, as shown in Fig. {\ref{fig_3}}(f), so that it cannot enter into the subsequent DLA stage \cite{hua2}. This indicates that with the increase of the self-generated azimuthal magnetic field, the electron injection process is gradually suppressed, which is the reason why most of the resonance electrons comes from the front surface of the plasma target, as shown in Fig. \ref{fig_2}(c).

In summary, the physical mechanism of electron injection in DLA regime is clarified for the first time. 
It is found that the ion wave, excited by the longitudinal charge-separation electric field, plays a crucial 
role on the electron injection process. The ion wave induces a localized longitudinal electric field, which 
acts as a set of potential wells and determines the injection path for the electrons from the channel edge. 
In addition, it is pointed out that the intense self-generated azimuthal magnetic field can suppress the 
injection process of electrons by diverting them towards the opposite side of the laser propagation direction. 
This mechanism governs the electron injection dynamics in a relativistic transparency regime.

This work is supported by the National Key Program for S\&T Research and Development, Grant No. 2016YFA0401100; the SSTDF, Grant No. JCYJ20160308093947132; the National Natural Science Foundation of China (NSFC), Grant Nos. 91230205, 11575031, 11575298, and 11705120. B. Q. acknowledges the support from Thousand Young Talents Program of China. K. J. would like to thank Z. Li, H. Z. Liu, X. B. Li and J. W. Liu for their useful discussions and help.

\end{document}